\documentclass[sigconf,nonacm]{acmart}

\usepackage{hyphenat}
\usepackage{xspace}

\usepackage{amsthm}
\theoremstyle{plain}

\newcommand{\textcite}[1]{\citeauthor{#1} \shortcite{#1}}

\newcommand{\hide}[1]{}

\makeatletter
\newcommand{\iffont}[2]{\ifthenelse{\equal{\f@family}{#1}}{#2}{}}
\makeatother

\iffont{ptm}{
  \usepackage{mathptmx}

  \DeclareSymbolFont{greek}{OML}{cmm}{m}{n}
  \DeclareMathSymbol{\alpha}{\mathalpha}{greek}{"0B}
  \DeclareMathSymbol{\beta}{\mathalpha}{greek}{"0C}
  \DeclareMathSymbol{\gamma}{\mathalpha}{greek}{"0D}
  \DeclareMathSymbol{\delta}{\mathalpha}{greek}{"0E}
  \DeclareMathSymbol{\epsilon}{\mathalpha}{greek}{"0F}
  \DeclareMathSymbol{\zeta}{\mathalpha}{greek}{"10}
  \DeclareMathSymbol{\eta}{\mathalpha}{greek}{"11}
  \DeclareMathSymbol{\theta}{\mathalpha}{greek}{"12}
  \DeclareMathSymbol{\iota}{\mathalpha}{greek}{"13}
  \DeclareMathSymbol{\kappa}{\mathalpha}{greek}{"14}
  \DeclareMathSymbol{\lambda}{\mathalpha}{greek}{"15}
  \DeclareMathSymbol{\mu}{\mathalpha}{greek}{"16}
  \DeclareMathSymbol{\nu}{\mathalpha}{greek}{"17}
  \DeclareMathSymbol{\xi}{\mathalpha}{greek}{"18}
  \DeclareMathSymbol{\pi}{\mathalpha}{greek}{"19}
  \DeclareMathSymbol{\rho}{\mathalpha}{greek}{"1A}
  \DeclareMathSymbol{\sigma}{\mathalpha}{greek}{"1B}
  \DeclareMathSymbol{\tau}{\mathalpha}{greek}{"1C}
  \DeclareMathSymbol{\upsilon}{\mathalpha}{greek}{"1D}
  \DeclareMathSymbol{\phi}{\mathalpha}{greek}{"1E}
  \DeclareMathSymbol{\chi}{\mathalpha}{greek}{"1F}
  \DeclareMathSymbol{\psi}{\mathalpha}{greek}{"20}
  \DeclareMathSymbol{\omega}{\mathalpha}{greek}{"21}
  \DeclareMathSymbol{\varepsilon}{\mathalpha}{greek}{"22}
  \DeclareMathSymbol{\vartheta}{\mathalpha}{greek}{"23}
  \DeclareMathSymbol{\varpi}{\mathalpha}{greek}{"24}
  \DeclareMathSymbol{\varrho}{\mathalpha}{greek}{"25}
  \DeclareMathSymbol{\varsigma}{\mathalpha}{greek}{"26}
  \DeclareMathSymbol{\varphi}{\mathalpha}{greek}{"27}
  \DeclareSymbolFont{otone}{OT1}{cmr}{m}{n}
  \DeclareMathSymbol{\Gamma}{\mathalpha}{otone}{0}
  \DeclareMathSymbol{\Delta}{\mathalpha}{otone}{1}
  \DeclareMathSymbol{\Theta}{\mathalpha}{otone}{2}
  \DeclareMathSymbol{\Lambda}{\mathalpha}{otone}{3}
  \DeclareMathSymbol{\Xi}{\mathalpha}{otone}{4}
  \DeclareMathSymbol{\Pi}{\mathalpha}{otone}{5}
  \DeclareMathSymbol{\Sigma}{\mathalpha}{otone}{6}
  \DeclareMathSymbol{\Upsilon}{\mathalpha}{otone}{7}
  \DeclareMathSymbol{\Phi}{\mathalpha}{otone}{8}
  \DeclareMathSymbol{\Psi}{\mathalpha}{otone}{9}
  \DeclareMathSymbol{\Omega}{\mathalpha}{otone}{10}
  \DeclareSymbolFont{syms}{OML}{cmm}{m}{it}
  \DeclareMathSymbol{\partial}{\mathord}{syms}{"40}
  \DeclareMathAlphabet{\mathbold}{OML}{cmm}{b}{it}
  \DeclareSymbolFont{largesymbols}{OMX}{cmex}{m}{n}

}
\usepackage{mathrsfs}

\usepackage{url}



\usepackage{subcaption}

\usepackage{tikz}
\usetikzlibrary{positioning, calc, shapes.geometric, shapes, shapes.multipart, arrows.meta, arrows, decorations.markings, external, trees}
\tikzset{
node distance=0.5cm, 
}
\tikzstyle{Arrow} = [
	thick, 
	decoration={
		markings,
		mark=at position 1 with {
			\arrow[thick]{latex}
			}
		}, 
	shorten >= 0.5pt, preaction = {decorate},
	]

\AtBeginDocument{%
  \providecommand\BibTeX{{
    \normalfont B\kern-0.5em{\scshape i\kern-0.25em b}\kern-0.8em\TeX}}}

\copyrightyear{2024}
\acmYear{2024}
\setcopyright{acmlicensed}\acmConference[]{}{}{}
\acmBooktitle{}
\acmPrice{15.00}
\acmDOI{}
\acmISBN{xxx-xxx-xxx-xxx-x/xx/xx}

\begin{document}

\title[The Effects of Enterprise Social Media on Communication Networks]{The Effects of  Enterprise Social Media\\on Communication Networks}


\author{Manoel Horta Ribeiro}\authornote{Work done while at Microsoft.}
\affiliation{%
  \institution{Princeton University}
  \country{Princeton, US}
}
\email{manoel@cs.princeton.edu}

\author{Teny Shapiro}
\affiliation{%
  \institution{Microsoft}
  \country{San Francisco, US}
}
\email{teshapi@microsoft.com}

\author{Siddharth Suri}
\affiliation{%
  \institution{Microsoft Research}
  \country{Redmond, US}
}
\email{suri@microsoft.com}

\renewcommand{\shortauthors}{Horta Ribeiro et al.}

\begin{abstract}
Enterprise social media platforms (ESMPs) are web-based platforms with standard social media functionality, e.g., communicating with others, posting links and files, liking content, etc., yet all users are part of the same company. 
The first contribution of this work is the use of a difference-in-differences analysis of $99$ companies to measure the causal impact of ESMPs on companies' communication networks across the full spectrum of communication technologies used within companies: email, instant messaging, and ESMPs.
Adoption caused companies' communication networks to grow denser and more well-connected by adding new, novel ties that often, but not exclusively, involve communication from one to many employees. Importantly, some new ties also bridge otherwise separate parts of the corporate communication network.
The second contribution of this work, utilizing data on Microsoft's own communication network, is understanding how these communication technologies connect people across the corporate hierarchy.
Compared to email and instant messaging, ESMPs excel at connecting nodes distant in the corporate hierarchy both vertically (between leaders and employees) and horizontally (between employees in similar roles but different sectors). 
Also, influence in ESMPs is more `democratic' than elsewhere, with high-influence nodes well-distributed across the corporate hierarchy.
Overall, our results suggest that ESMPs boost information flow within companies and increase employees' attention to what is happening outside their immediate working group above and beyond email and instant messaging.
\end{abstract}

\ccsdesc[500]{Human-centered computing~Empirical studies in collaborative and social computing}
\keywords{enterprise social media, observational studies, network analysis}

\maketitle

\newpage 

\section{Introduction}

Enterprise social media platforms (ESMPs) are web-based platforms where employees within the same company interact with coworkers and broadcast messages~\cite{leonardi2013enterprise}.
ESMPs can be traced back to the 2000s, e.g., IBM's Beehive~\cite{dimicco2008motivations} and HP's Watercooler~\cite{brzozowski2009watercooler}.
Previous work describing these early systems suggests they may strengthen corporate communication by increasing employees' attention to what is happening outside of their immediate working group~\cite{brzozowski2009watercooler};
or by helping to form communities of practice that can boost organizational performance~\cite{lesser2001communities} and the development of employees' social capital~\cite{leidner2010assimilating}.
This work closely relates to other research on the effect of connectivity in the workplace. Individuals in better-connected social settings relay information more efficiently~\cite{reagans2003network}, and are more trusting~\cite{raub1990reputation}, productive~\cite{reagans2001networks}, and creative~\cite{soda2021networks}.

ESMPs have gathered increased interest with the growth of remote work and its associated challenges~\cite{sivunen2023role}.
While remote work is attractive to many due to the flexibility it affords~\cite{smite2023work}, 
it can make communication within companies siloed~\cite{yang2022effects}, 
undermine the employee onboarding process~\cite{yu2023large}, 
and curb innovation~\cite{brucks2022virtual}.
ESMPs, in that context, can potentially address these challenges by creating new information flows and opportunities to connect.

\begin{figure*}
\centering
    \includegraphics[width=0.9\textwidth]{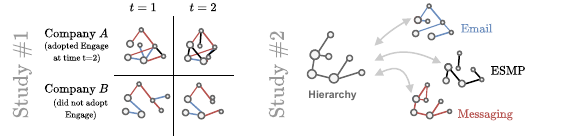}
    \caption{
    Study \#1: With a difference in differences analysis of 99 companies that suddenly adopted Engage, we study the impact of ESMPs on companies' communication networks across technologies (Instant Messaging, Email, ESMPs).
    We compare the communication networks of companies that adopted (Company A) or did not adopt (Company B) Engage. Then, we measure the effect of Engage by comparing network properties of the communication network before ($t_1$) and after ($t_2$) some of the companies adopted it.
    Study \#2: Analyzing Microsoft's communication network, we characterize how  ESMPs differ from email and instant messaging in bridging the company's hierarchy (\textit{right}). Modeling Microsoft's corporate hierarchy as a tree, we study how connections in different communication technologies bridge different levels and sectors of the company.}
    \label{fig:summary}
\end{figure*}

The fundamental questions this work addresses are: Do ESMPs create new connections and information flows within companies? Are these connections different from those established in other communication technologies like email or instant messaging? 
We report findings on two observational studies, depicted in Fig.~\ref{fig:summary}.
Specifically, we focus on the relationship between ESMPs and companies' communication networks. Nodes in this network represent employees, whereas edges represent information being sent (e.g., an email). These networks are built considering three communication technologies associated with the Microsoft ecosystem: ESMPs (Engage), email (Outlook), and Instant Messaging (Teams).

We studied 99 companies that suddenly adopted Engage, a popular Enterprise Social Media platform by Microsoft (formerly Yammer). Using a difference in differences (DiD) design, we measure the causal impact of ESMPs on companies' communication networks, considering metrics that range from the number of ties to their embeddedness, i.e., the number of shared connections between any two ties.
With this data, however, we cannot understand who is being connected by Engage within the company's organizational hierarchy. Thus, we additionally conducted an in-depth cross-sectional case study of Microsoft, analyzing its communication network and organizational hierarchy.
We compare how ESMPs and other communication technologies bridge the corporate hierarchy and explore the relationship between centrality in the communication network and its importance in the corporate hierarchy. The two studies are visually summarized in Fig.~\ref{fig:summary}.

\vspace{.5mm}
\noindent
\textbf{Findings.}
We present three key results.
First, we find that adopting ESMPs causes more new, non-redundant edges to be added to a company's communication network.
This is supported by our difference in differences analysis, where we find that new ties are added to companies adopting Engage (7.8\% average increase in the 20 weeks following adoption; $p<0.01$) and that this increase in ties is not explained by ``pre-existing'' ties, but rather by a large increase in the number of new connections happening between employees.
Second, adopting ESMPs causes weak ties to be added to companies' communication networks.
In the difference in differences analysis, we find that the average tie strength in companies' communication networks decreases (e.g., the embeddedness of the ties drops by 3.5 percentage points in the 20 weeks following adoption; $p<0.01$). Further, the Microsoft case study suggests that ESMPs' weak ties bridge the company's hierarchy across sectors and levels.
Third, adopting ESMPs causes a few employees to become more central in the company's communication network.
We calculate the PageRank distribution across organizations weekly and find that they become more unequal following the adoption of Engage.
In that direction, the Microsoft case study suggests that Engage ``democratizes'' influence within a company's communication network since highly central employees seem to be better distributed across the company's hierarchy in ESMPs than in other communication technologies.

\vspace{.5mm}
\noindent
\textbf{Implications.}
We provide empirical, quasi-experimental evidence that ESMPs boost information flow within companies and increase employees' attention to what is happening outside their immediate working group.
Our results further indicate that ESMPs may help counteract some downsides of companies  ``going remote,'' like the vanishing of weak ties and the siloization of communication~\cite{yang2022effects}.

\vspace{.5mm}
\noindent
\textbf{Ethics and Privacy.}
All data was de-identified and analyzed in aggregate, and the researchers viewed no individual-level data. Data use was compliant with Microsoft products' terms of service.
This study was approved by an IRB (details to be given in the final version to maintain anonymity). This work is the result of a collaboration between industry and academia.
This study was approved by the Microsoft IRB under approvals 10648 and 10653.

\section{Background and Related work}

We build upon two bodies of work: research on social technologies in corporate environments and on connectivity in the workplace.

\begin{figure*}[t]
    \centering
    \includegraphics[width=0.9\textwidth]{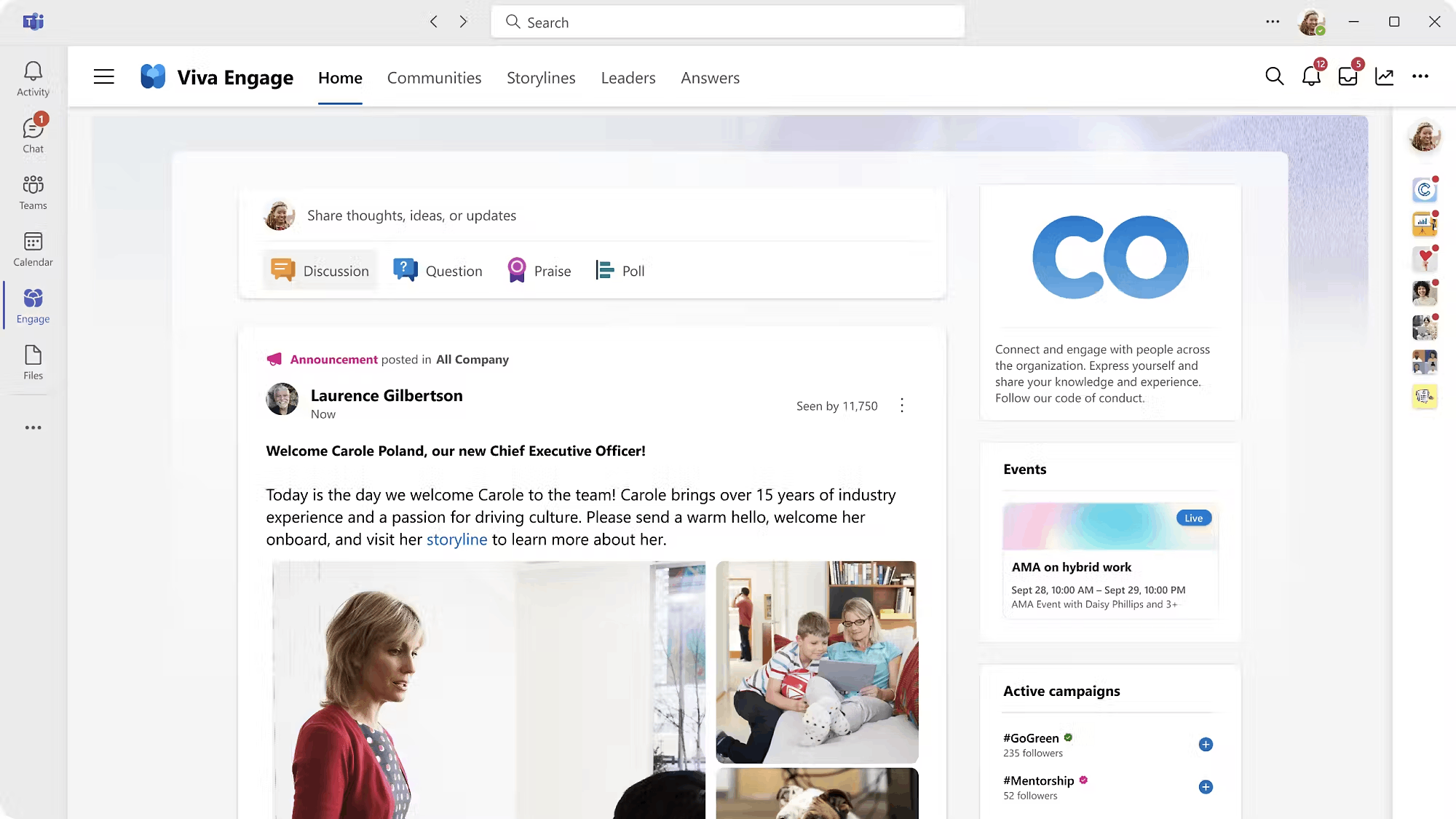}
    \caption{Engage, the ESMP studied here. Engage offers standard social media functionality (algorithmic feeds, posts, reactions) and is integrated within Microsoft Teams, which offers instant messaging functionalities.}
    \label{fig:enter-label}
\end{figure*}

\vspace{0.5mm}
\noindent
\textbf{Enterprise social media.}
With the advent of Web 2.0, emphasizing user-generated content and participatory culture, major companies developed and used various social technologies~\cite{mcafee2006enterprise}.
These ranged from generic software allowing employees to easily deploy blogs on the intranet~\cite{yardi2008pulse, efimova2007crossing} to tailor-made platforms aggregating social media functionality with the organization's employee directory. 
In that context, early papers provide a detailed description of the usage, deployment, and design of early ESMPs~\cite{brzozowski2009watercooler, dimicco2008motivations, efimova2007crossing, kolari2007structure}. 
For example, the WaterCooler system deployed within HP in the late 2000s is described in \citet{brzozowski2009watercooler}. The system allowed employees to create profiles and tag themselves and each other. It also allowed employees 
to write posts and search them with sophisticated filters.
Also, in a similar spirit,  \citet{dimicco2008motivations} describes the social network deployed at IBM around 2008. ``Beehive'' allowed personal and professional sharing---employees could create and share lists and albums, as well as comment on other employees' profiles.

Other work has focused on understanding the motivations behind ESMP use. 
For example, past research found that users' participation in ESMPs was correlated to the involvement of recent managers or coworkers and with receiving feedback, e.g., in the form of comments~\cite{brzozowski2009effects}.
Indeed, employees often described being frustrated when they contributed to blogs, but their expectations of attention were unmet~\cite{yardi2009blogging}. 
It is worth noting that the adoption of social media at work was not without its hiccups. Early adopters of `employee blogs' (the proto-ESMPs) had to navigate obscure guidelines of what was and what was not allowed~\cite{efimova2009passion}---and were sometimes fired for behavior seen as inappropriate, e.g., discussing everyday life at work~\cite{coneRiseBlog2005}. In her book, ``Passion at Work,'' \citet{efimova2009passion} argues employee bloggers flourished (even amidst these setbacks) because blogging creates valuable connections and helps to improve ideas by having to articulate them in public.

Perhaps assuming their inevitable adoption, other research has examined the correlation between ESMP use and a variety of outcomes of interest to companies.
For example, \citet{aboelmaged2018knowledge} suggests that ESMP usage is correlated with employees' productivity;
\citet{pitafi2018investigating} link ESMPs to employees' agility, i.e., their ability to respond to market changes.
\citet{luo2018can} associate ESMP use to affective organization commitment
Yet, evidence of whether and how ESMPs impact companies remains preliminary. 
Relevant research relies on observational cross-sectional designs (often survey-based) that are not able to disentangle confounders associated with both ESMP use and outcomes of interest or fails to consider the full spectrum of communication technologies within a company~\cite{azaizah2018impact}.
Both these limitations threaten the validity of these papers' results. 
On the one hand, ESMP use, as these papers study, is largely impacted by self-selection, and the studies may be measuring characteristics that lead users to use ESMPs.
In other words, it may be that more productive employees adopt ESMPs, not that ESMPs make them more productive.
On the other hand, these studies often consider ESMPs in isolation and ignore that they may replace rather than complement existing communication mediums within the company.
Here, we address these limitations by employing a credible identification strategy to isolate the causal effect of ESMPs on company-level outcomes and by considering the full spectrum of corporate communication technologies. 

\vspace{0.5mm}
\noindent
\textbf{Connectivity on the workplace.}
The field of network science has found that the social structure is consequential in explaining outcomes associated with individuals, companies, and even entire economies~\cite{watts1999small,barabasi2014linked}. 
In that context, a vast, interdisciplinary literature has described and tried to understand the impact of social structure and connectivity in the workplace~\cite{reagans2003network,raub1990reputation,soda2021networks}.
For example, \citet{granovetter1973strength} showed how `weak ties,' casual connections, and loose acquaintances play a disproportionally large role for job-seekers.
\citet{coleman1988social} highlighted the importance of network closure, noting that tightly connected networks increase cooperation and social capital.
\citet{burt_structural_1992} proposed the concept of `structural holes,' arguing that individuals positioned as mediators of two closely connected groups  (often referred to as bridges) gain important competitive advantages within companies (and the most varied scenarios).

Much of the foundational research on connectivity in the workplace (and in general) happened before the popularization of the Web and the widespread adoption of Web-based communication platforms at work~\cite{coleman1988social,granovetter1973strength,burt_structural_1992}. In that context, online communication networks provided a suitable context for systematic, empiric investigation of these theories.
A study of 21 billion Facebook friendships found that connectedness among individuals with varying economic status is critical to upward income mobility~\cite{chetty2022social}.
A large experiment on LinkedIn manipulated the strength ties recommended by the website's recommendation algorithm, finding experimental causal evidence
supporting the `strength of weak ties'~\cite{rajkumar2022causal}.
An extensive analysis of a dynamic (email) social network of a university~\cite{kossinets2006empirical} examined, among other things, the temporal stability of bridges and structural holes, finding them to be largely unstable.
This work draws from these large-scale empirical investigations as we attempt to uncover whether mechanisms idealized by the designers of ESMPs happen in practice.

Concerns about how connectivity impacts the workplace were heightened by the advent of remote work, drastically accelerated by the COVID-19 pandemic~\cite{phillips2020working}.
Recent work has indicated that remote work trends may harm companies by inadvertently altering their internal social structure.
\citet{yang2022effects} found that firm-wide remote work led to more static and siloed collaboration networks.
\citet{yu2023large} found significant gaps in the social networks of workers onboarded remotely during the pandemic, indicating that they may be at a professional disadvantage.
\citet{brown2023effects} found that time spent alone during the pandemic correlates with poorer organizational well-being.
Here, we investigate to which extent ESMPs may counteract some of the challenges associated with transitioning to remote work.

\vspace{0.5mm}
\noindent
\textbf{Engage.} In Figure~\ref{fig:enter-label}, we depict Engage, the ESMP considered in this study. Engage offers features like stories and posts that mirror familiar social media experiences (e.g., Facebook, Twitter) and integrates seamlessly with Microsoft Teams and other Microsoft 365 tools (which we also consider here when looking at companies' communication networks).

\begin{figure}[t]
    \centering
    \includegraphics[width=\linewidth]{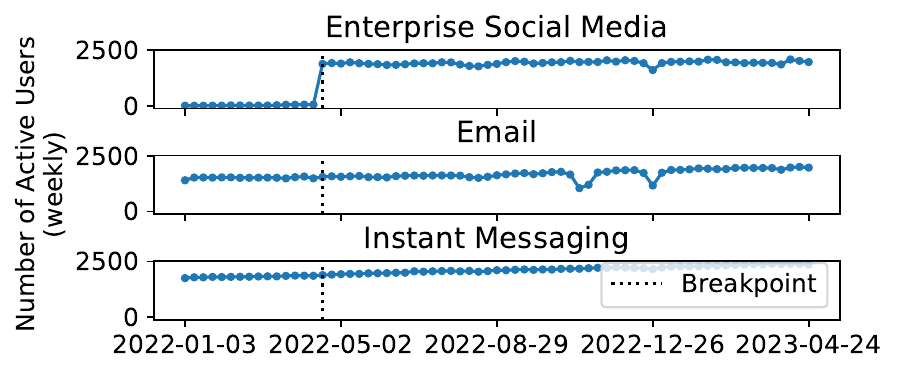}
    \caption{For a random company in our sample, we show the number of active users across ESM, Email, and Instant Messaging in the 69-week study period considered. A vertical dotted line indicates when the company suddenly adopted Engage, Microsoft's ESMP solution. We use the discontinuity around the adoption of Engage to study the effect of its adoption on the company's communication network (see Sec.~\ref{sec:did} for methodological details).}
    \label{fig:breakingpoint_example}
\end{figure}

\section{Study \#1: Impact of Engage Adoption}

\subsection{Data}
\label{sec:data1}

For Study \#1, we consider all companies that suddenly adopted Engage between 2022-01-03 and 2023-05-01; see Sec.~\ref{sec:selection_companies} for their selection. 
We refer to this dataset as the `adoption data.'
We consider all corporate communication happening within Microsoft's products, specifically, all that was written and read across Outlook (email), Teams (instant messaging), and Engage (Microsoft's ESMP solution; here, we also consider reactions, e.g., likes).

We use these traces to create (directed) communication networks. Each employee in the company is a node in the network, and edges indicate information flow; e.g., an edge from $u$ to $v$ indicates that $v$ has consumed information produced by $u$.
We consider three different kinds of edges for each data source. 
For email and instant messaging, edge types capture the number of participants involved in a conversation: 
\textit{one-on-one} or \textit{one-to-many}.
For ESMPs, edges capture when employees \textit{view}, \textit{reply}, or \textit{react} to someone's content.

We create communication networks at a weekly granularity, i.e., for each company, and each week, we create a graph capturing all communication happening that week. 

\begin{figure*}
    \centering
    \includegraphics[width=\linewidth]{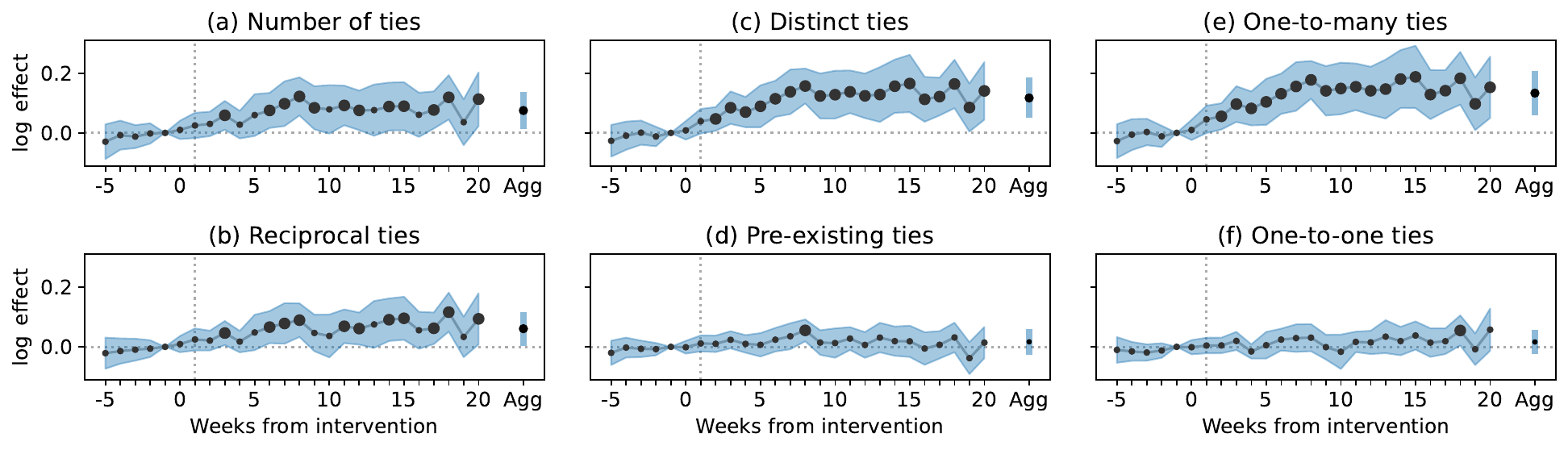}
    \caption{
    Results of our difference in differences analysis. 
    On the $x$-axis, we show the number of weeks from the intervention, ranging from -5 to 20. 
    Note that interventions are enacted on Week 1 and that Week $-$1 is the reference point.
    On the $y$-axis, we show the $\log$ effect, i.e., the effect on the outcome (one per panel). 
    The shaded area corresponds to 95\% confidence intervals. 
    On the right-hand side of each plot, we also show the aggregated effect (`Agg'), i.e., the average weekly effect across weeks 1 to 20.
    }
    \label{fig:dd1}
\end{figure*}

\subsection{Selecting Companies Adopting VE}
\label{sec:selection_companies}

To measure the causal effect of ESMPs, we obtain companies that \textit{suddenly} adopted Engage (VE), Microsoft's ESMP solution.
VE is sold as part of the Microsoft 365 Enterprise offering, an all-in-one solution for companies that include Teams (instant messaging), Outlook (e-mail), as well as other software (e.g., Word, Powerpoint, Excel).
However, Engage is substantially more complex to set up than other communication software included. 
Companies must set up initial communities, leaders must populate their profiles, and community leaders must be appointed to moderate communities. 
Thus, adoption here refers not to the purchasing of software but to the corporate decision to kickstart Engage as one of the internal means of communication.

In some cases, ESMP adoption may be slow, e.g., companies may onboard different departments one by one; in others, this adoption is sudden, with a substantial portion (if not all) of the company being introduced to the ESMP simultaneously. 
Here, we study the impact of adopting Engage by using differences in differences, a quasi-experimental design that requires suddenness, and therefore, we use simple heuristics to select a subset of companies that did so. Specifically, we obtain a sample of $n=99$ by following the four steps described below.

\begin{enumerate}
\item 
We obtain all companies using Engage worldwide between 2022 Q1 and 2023 Q2 (inclusive) that experienced a 50\% growth in the number of Engage users and that had at least 1000 Engage users at one of the quarters considered. 

\item 
Considering these companies, for each of the 69 weeks between 2022-01-03 and 2023-05-30, we estimate the percentage of employees using Engage (out of the number of weekly users using other Microsoft 365 products). We refer to this quantity as $y_{c,t}$, where $c \in \{1,\ldots,99\}$ indicates the company and $t \in \{1,\ldots,69\}$ indicates the week.

\item 
For each company $c$ associated with an Engage-usage time-series $y_c = \{y_{c,1}, \ldots, y_{c, 69}\}$, we find the breakpoint $i$ solving
\[
    \min_i \sum_{t < i} (y_{c,t} - \overline{y}_{c, i < t})^2  + \sum_{t \geq i} (y_{c,t} - \overline{y}_{c, i \geq t})^2, 
\]
where $\overline{y}_{c, i < t}$ is the mean of  $\{y_{c,t}\}_{t<i}$. Note that this method (following \citet{truong2020selective}) always yields a breakpoint.

\item 
Finally, we filter companies, keeping only those in which the time series corresponds to a sharp discontinuity. Specifically, we consider only companies whose 
i)~breaking point is not in the first ten weeks of the study period;
ii)~ESMP adoption before the breaking point was at most than 7\%;
iii)~ESMP adoption after the breaking point was at least than 15\%. 
\end{enumerate}

Empirically, this approach lets us obtain companies that suddenly adopted Engage with high precision. 
We illustrate the results showing the (absolute) number of daily active users across platforms for a random company in our sample in Fig.~\ref{fig:breakingpoint_example}. Note that the figure does not convey information about companies' communication networks, just about daily usage of products.

\subsection{Difference in differences}
\label{sec:did}

To study the causal effect of adopting ESMPs, we use a difference in differences approach.
We compare `treated' units (\textit{companies} who adopted Engage) with `control units' (\textit{companies} who did not) under the parallel trends assumption, i.e., that in the absence of treatment, the difference between the `treatment' and `control' groups remains constant.

Considering, for example, the number of ties in a company's weekly information network ($Y$), the simple DiD estimator is 
\begin{equation}
\label{eq:dd1}
    \hat{\delta}_{DD} = (\bar{Y}_{\text{Post},\text{Treatment}} - \bar{Y}_{\text{Pre},\text{Treatment}}) - (\bar{Y}_{\text{Post},\text{Control}} - \bar{Y}_{\text{Pre},\text{Control}}),
\end{equation}
where the notation $\bar{Y}_{A, B}$ denotes the average outcome across units in group $B \in \{\text{Treatment}, \text{Control}\}$ in period $A \in \{\text{Pre}, \text{Post}\}$. 
Intuitively, this estimator creates a counterfactual estimate of how the treatment group would have progressed without treatment and compares it with the observed change in the outcome. 
It can be shown that $\hat{\delta}_{DD}$ estimates the average treatment effect on the treated (ATT), given that the parallel trends assumption holds~\cite{sun2021estimating}. 
Considering the number of ties as the outcome, we have that, under the parallel trends assumption,  $\hat{\delta}_{DD}$ is the increase in the number of ties caused by the adoption of Engage.

Our analysis of the adoption data, however, is further complicated by the fact that 1) all companies are eventually treated, i.e., all companies in our sample eventually adopt Engage, and 2) each company has a distinct treatment time.
Fortunately, recent work has developed more complex DiD estimators that accommodate this setup.
Here, we use the methodology proposed by Sun and Abraham~\cite{sun2021estimating}. 

The ``leads and lags'' specification considers a panel of $i=1,\ldots, N$ units for $t=1,\ldots, T$ calendar periods and estimates dynamic coefficients $ \delta_{i,\ell}$ (where $\ell$ is the relative time until unit $i$ receives the treatment). 
These can be aggregated to estimate the effect of the treatment at different relative times, e.g., 1 week after the intervention. We estimate these coefficients using a two-way fixed effect regression of the form
\begin{equation}
\label{eq:dd2}
    Y_{i,t} = \alpha_i + \lambda_t + 
    \sum_{\ell < -2} \delta_{i,\ell}  D^\ell_{i,t} + 
    \sum_{\ell \geq 0} \delta_{i,\ell}  D^\ell_{i,t} + 
    \epsilon_{i,t}.
\end{equation}
Here, $Y_{i,t}$ is an outcome associated with unit $i$ on time $t$, $\alpha_i$ and $\lambda_t$ are unit and time fixed effects, $D^\ell_{i,t}$ is an indicator variable for unit $i$ being $\ell$ periods away from when it receives treatment.

Intuitively, this approach compares treated units with ``yet-to-be'' treated units, e.g., comparing companies that adopted Engage with companies that are yet to do so. However, recent work has shown that the estimator is biased because, under the hood, there are `forbidden comparisons' between pairs when both have already been treated. To address that, Sun and Abraham~\cite{sun2021estimating} modify the estimator procedure, preventing the comparison of such units. 
Nonetheless, we obtain coefficients equivalent to those depicted in Eq.~\ref{eq:dd2} that allow us to estimate the effect of treatment $\{\delta_{i,\ell}\}_{l>0}$ and to assess the parallel trends assumption  $\{\delta_{i,\ell}\}_{l<-2}$.

\subsection{Results}

\begin{figure}
    \centering
    \includegraphics[width=\linewidth]{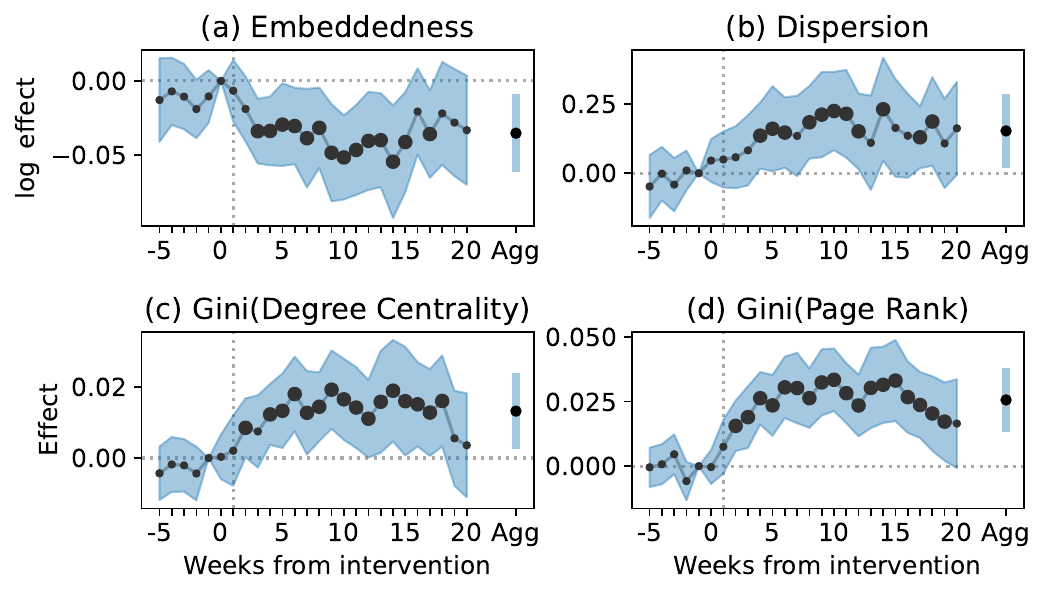}
    \caption{
    Effects of adopting Engage on metrics associated with tie strength (a-b) and centrality (c-d). 
    Again, on the $x$-axis, we show the number of weeks from the intervention, ranging from -5 to 20; we also show the aggregated effect (`Agg').
    Note that outcomes here are not log-transformed -- they should be interpreted as percentage point changes.
    }
    \label{fig:dd2}
\end{figure}

To understand the effects of ESMP adoption on companies' communication networks, we conduct a difference in differences regression analysis of $99$ companies that suddenly adopted Engage, Microsoft's ESMP solution, throughout 69 weeks.
We consider a variety of outcomes (e.g., number of ties) associated with the weekly communication networks of each company.
Note that we log-transform several of the outcomes, allowing us to interpret the effects multiplicatively and better accommodate the differences in size between the companies considered.%
\footnote{A coefficient  $\delta$ in a linear regression with a log-transformed outcome corresponds to a multiplicative factor $e^\delta$; e.g., $\delta=0.09$ corresponds to a relative increase by 9.4\%}
We report detailed statistics for the difference in difference analysis, as well as additional plots in Table~\ref{tab:app}; results are summarized in Fig.~\ref{fig:dd1} and Fig.~\ref{fig:dd2}.
When reporting effects in the main text, we consider the average weekly effect across the 20 weeks after the adoption of ESMPS.

\vspace{.5mm}
\noindent
\textbf{Number of ties.}
Part of the promise of enterprise social media is to strengthen companies' communication networks in general, yet their adoption could lead employees to merely \textit{substitute} one means of communication by another, e.g., they could use instant messaging or emails less often so that their ESMP usage did not increase the total information flow.
Using our DiD framework, we analyze the effect of adopting Engage on the number of weekly ties within companies' communication networks. Note that this corresponds to the sum of the weights of the edges in the communication networks.
We find a significant increase of 7.8\% in the number of ties across all three communication networks considered. 
In Fig.~\ref{fig:dd1}a, we show how this effect is spread somewhat consistently through the weeks after the adoption. 
Additionally, note that we find that there is no difference between treatment and (yet-to-be-treated) control groups in the six weeks before the intervention ($-$5 to 0), which suggests that the parallel trends assumption holds.

\vspace{.5mm}
\noindent
\textbf{Distinct vs. pre-existing ties.}
While we observe a significant increase in the number of total ties, it could be that Engage either 
1) increases the number of distinct connections employees have or 
2) strengthens their already existing connections. 
In that context, we use the same DiD framework considering 
1) the number of edges in the unweighted communication network (which we refer to as ``distinct ties'');
2) the sum of the edges in the weighted communication network, considering only edges between pairs of nodes that exchanged one message in the first five weeks of the study period (``pre-existing ties'').
Results are shown in Fig.~\ref{fig:dd1}c--d. 
The increase in distinct ties is statistically significant and even stronger than the effect of ties in general (12.4\%). 
In contrast, the increase in pre-existing ties is much weaker and not statistically significant (1.7\%). 
This indicates that ESMP adoption impacts companies' communication networks mostly through new, novel ties.

\vspace{.5mm}
\noindent
\textbf{One-to-many vs. one-to-one ties.}  
To understand whether the increase in ties comes from broadcasted (one-to-many) or unicasted (one-to-one) information, we run the difference in difference regression considering exclusively: 
1) one-to-one messages on instant messaging and emails and replies/reactions on ESM; and 
2) one-to-many messages on instant messaging and email and views on ESM.
Results are shown in Fig.~\ref{fig:dd1}e--f. 
Similar to the distinct vs. pre-existing scenario, we find that the effect for one-to-many ties is statistically significant and even stronger than that of ties in general (14.3\%), while the effect for one-to-one ties is much weaker and not statistically significant (1.6\%). This indicates that ESMPs increase information flow within companies through new, novel, one-to-many ties, following the broadcasting nature of social media.

\vspace{.5mm}
\noindent
\textbf{Reciprocal ties.} 
The previous results show adopting ESMPs create additional information flow between previously unconnected users through broadcasts --- but are these connections reciprocal or unidirectional?
For example, maybe Engage is a new communication channel for leaders to broadcast information to their subordinates.
We explore whether that is the case by measuring the effect of Engage adoption on the number of distinct reciprocal ties in the network (see Fig.~\ref{fig:dd1}b). 
We find a significant increase of 6.2\%, which suggests that bi-directional connections are also formed through the network.

\vspace{.5mm}
\noindent
\textbf{Tie strength.} 
Communication networks like the ones we consider here capture interpersonal relationships of varied `strengths.' 
Strong ties involving extensive interaction are usually `embedded' in the network; the two individuals involved in the connection typically have many other connections in common, whereas weak ties involve few mutual connections\cite{granovetter1973strength}. 
To study how the adoption of Engage impacted tie strength, we analyze the evolution of two metrics associated with tie strength: embeddedness and dispersion. 
For a given tie between two employees, embeddedness is the number of nodes connected to both~\cite{marsden1984measuring}.
Considering these sets of `mutual relationships,' the dispersion captures the extent to which these individuals are not themselves well connected~\cite{backstrom2014romantic}.
Previous work indicates weak ties are well embedded (i.e., high embeddedness) and not dispersed (i.e., low dispersion).
Here, we consider normalized versions of both metrics; e.g., embeddedness is normalized by the possible number of shared neighbors between two nodes.

For each communication network in our analysis (one per week per company), we sampled $10{,}000$ edges at random and computed their embeddedness and dispersion. 
Note that outcomes here are not log-transformed; they are fractions between 0 and 1, which allow us to interpret results as percentage points increase (\textit{pp}).
Using a difference in differences regression, we find that adopting Engage significantly decreased embeddedness by 3.5 \textit{pp} and increased dispersion by 12.5 \textit{pp}; see Fig.~\ref{fig:dd2}a-b. 
This suggests that the ties added by Engage adoption are weak ties, bridging parts of the network that are not well connected. Had these ties just densified neighborhoods of the network that were previously well connected, we would expect the opposite; that tie strength would have increased.

\vspace{.5mm}
\noindent
\textbf{Centrality.} 
Individuals within communication networks may be more or less influential (or `central') depending on how much information they share and with whom they share information. In online social networks, few users are responsible for a large fraction of connections within the platform~\cite{mislove2007measurement}; thus, Engage may make fewer nodes more `central.'
Here, we consider two metrics of centrality: degree centrality, i.e., how many ties each employee has), and PageRank centrality, which also considers the importance of whom you are connected to. Note that we invert the edge direction when computing PageRank, as influence flows from who consumes information to whose information is consumed (but edges were created in the other direction).
Then, considering the centrality distributions, we calculate the Gini coefficient, capturing how concentrated in few nodes these measurements are.

We find significant increases in the Gini coefficient of both degree (1.3 \textit{pp}) and PageRank centrality distributions (2.6 \textit{pp}).
Analyzing indegree and outdegree centrality separately (see Fig.~\ref{fig:app1}), we find that individuals with high outdegree centrality mostly explain the effect observed for the degree centrality. 
Overall, these findings indicate that additional ties created by the adoption of Engage are concentrated in the hands of a few employees, who start to exert additional influence within the company. 

\begin{figure}
    \centering
    \includegraphics[width=\linewidth]{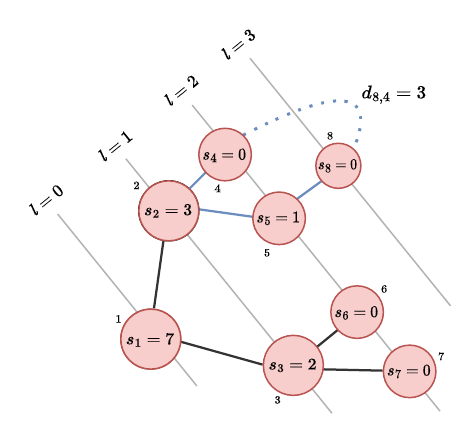}
    \caption{
    We show a toy hierarchical network with eight nodes (numbered) and key hierarchy-related metrics.
    For each node, we calculate:
    1)~the number of descendants $s$ in the tree, e.g., node \#$2$ has three descendants, \#$4$, \#$5$, and \#$8$, thus $s_2 = 3$;
    and
    2)~the distance $l$ from the tree's root to the node, e.g., node \#$6$ is two hops away from the CEO, thus $l_6 = 2$ (also known as the node's level).
    Last, we consider the distance $d$ of each node pair within the tree, e.g., the shortest path between nodes \#$4$ and \#$8$ is $\{(8,5), (5,2), (2,4)\}$, thus $d_{8,4} = 3$.
    }
    \label{fig:hie}
\end{figure}

\section{Study \#2: Microsoft Case Study}

\subsection{Data}

For Study \#2, we consider all corporate communication traces among Microsoft employees between 2022-06-17 and 2023-01-30.
We join this data with Microsoft's corporate hierarchy; see Sec.~\ref{sec:hierarchy} for details.
We refer to this dataset as `Microsoft's data.'
Again, we consider all corporate communication happening within Microsoft's products and construct communication networks similar to Study \#1 (see Sec.~\ref{sec:data1}).

We create one communication network for the entire study period.
The rationale here is that we use the adoption data for a longitudinal analysis, where we compare communication networks before and after the adoption of ESM. 
In contrast, for Microsoft's data, the cross-sectional analysis focuses on the relationship between corporate hierarchy and communication networks.

\subsection{Microsoft's hierarchy}
\label{sec:hierarchy}

For the Microsoft employee analysis, we obtained snapshots of the company's hierarchy between 2022-04-25 and 2023-01-16,%
\footnote{Snapshots are weekly, except for the week of 2022-05-30.}
and construct a tree $T_s$ such that, for each snapshot $s$,  $\langle u,v\rangle \in T_s$ if and only if $v$ directly reports to $u$. 
Through this process, we considered only the largest connected component in the original hierarchy, discarding 2.6\% (on average, never more than 3\%) of employees whose hierarchical structure was incomplete in the snapshot.

For each weekly graph, we calculate two metrics capturing the seniority and relative importance of workers within the company. 
First, we consider the number of (indirect) \textit{subordinates}, i.e., the number of workers that are below that individual in the hierarchy. 
Second, we consider each individual's \textit{level}; in the hierarchy, i.e., the distance of each individual to the tree's root (the CEO).
Last, we calculate the \textit{distance} in the hierarchy between all pairs of individuals, i.e., the minimum path length in $G_s^*$, the undirected version of $G_s$.
We aggregate these metrics across all snapshots by taking their average across the graphs associated with each snapshot.
Note that the first two metrics, \textit{level} and \textit{subordinates}, capture vertical distance in the company hierarchy --- individuals that are more influential within an organization tend to be closer to the CEO and have more subordinates.
In contrast, the third metric, \textit{distance}, also captures how far individuals are ``horizontally,'' e.g., individuals at similar roles in different branches.
We illustrate these three metrics in Fig.~\ref{fig:hie}.

We stress that analyzing the data of Microsoft's communication network along with the company's hierarchy allows us to derive rich insights that are impossible to obtain in the quasi-experimental setup of Study \#1. There, we know little about whom the ties added by Engage are connecting, as data are anonymized and we do not have companies' corporate structure. In contrast, here we can understand how different communication technologies (including ESMPs) may bridge the company both ``vertically'' (across levels in the hierarchy) and ``horizontally'' (across sectors of the company).

\subsection{Results}

Considering Microsoft data, we study ESMPs with the two other communication technologies used --- instant messaging and email --- considering metrics that capture the relationship between employees and the company's hierarchy.

\begin{figure}
    \centering
    \includegraphics[width=\linewidth]{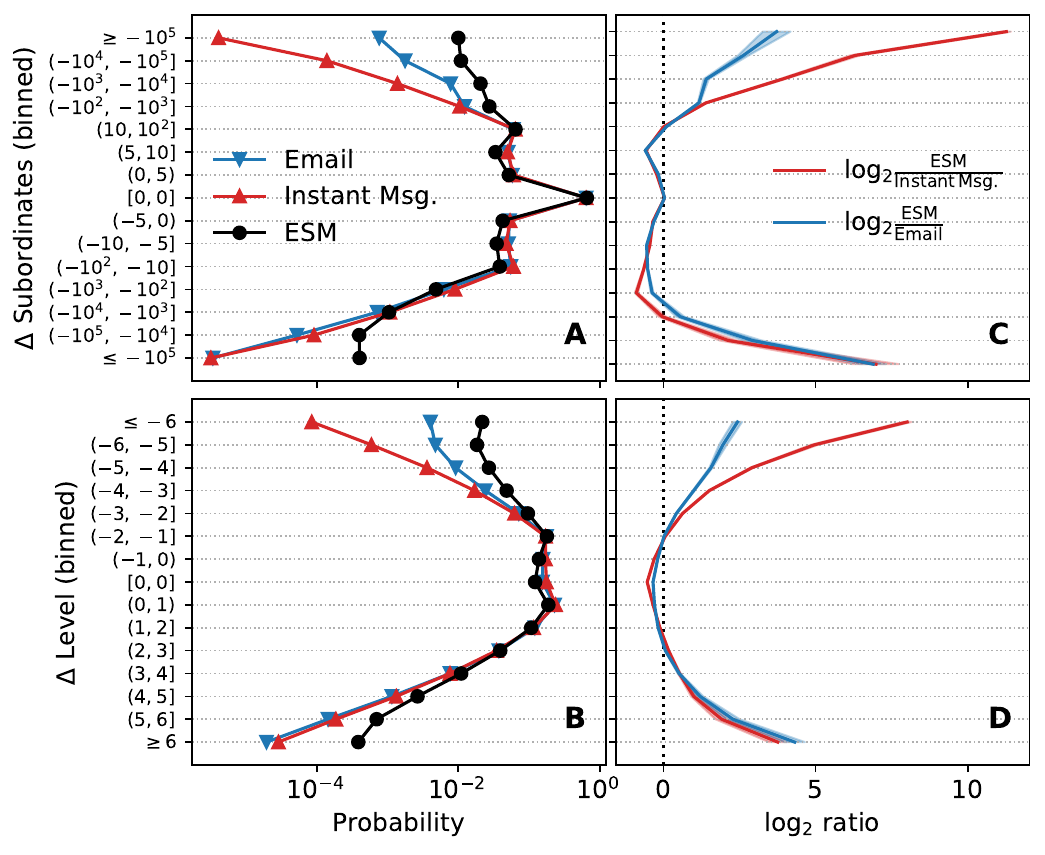}
    \caption{We show how ties in Microsoft's communication network bridge the corporate hierarchy vertically, computing,
    for each tie, the difference in subordinates and level for the employees involved.
    In A--B, show the distribution for these metrics considering ties that belong to each communication technology.
    In C--D we show the $log_2$ ratio of the distributions; we divide the fraction of ties occurring in each bin for ESMPs by the fraction of ties occurring in the same bin considering email (blue) and instant messaging (red) and then take the $\log_2$; Note that a $\log_2$ ratio $k$ means ties of that kind (e.g.  $\Delta\text{Level} \leq -6$), in the ESMPs are $2^k$ as frequent. We plot 95\% bootstrapped confidence intervals for the  $log_2$ ratios.
    }
    \label{fig:msft1}
\end{figure}

\vspace{.5mm}
\noindent
\textbf{``Vertical'' ties across the corporate hierarchy.}
Given a directed edge in the communication network, let $\Delta\text{Subordinates}$ be the number of subordinates of the sender minus the number of subordinates of the recipient, and the $\Delta\text{Level}$ be the level of the sender minus the level of the recipient.
We use these metrics to characterize how ties in Microsoft's aggregated communication network bridge the company's hierarchy vertically.
In Fig.~\ref{fig:msft1} A--B, we show the distribution of $\Delta\text{Level}$ and $\Delta\text{Subordinates}$ for each communication technology. 
In ESM, a higher fraction of edges connects individuals that are far away from each other in the network hierarchy.

To get a sense of the magnitude of these differences, we estimate the $\log_2$ ratio between the fraction of edges in ESM and other communication technologies for ties with different $\Delta\text{Level}$ and $\Delta\text{Subordinate}$ (Fig.~\ref{fig:msft1} C--D). A $\log_2$ ratio of $k$ means ties of a certain kind (e.g.,  $\Delta\text{Level} \leq -6$) are $2^k$ more frequent in the ESM than on another communication technology. 
We find that ESM ties are much more likely to connect individuals ``vertically''  far in the corporate ladder, with more messages being sent from the top to the bottom and vice versa. E.g., ties with $\Delta\text{Level}\geq6$ are 30 times more frequent in ESM than instant messaging $\log_2$ ratio $=8$).

\begin{figure}
    \centering
    \includegraphics[width=\linewidth]{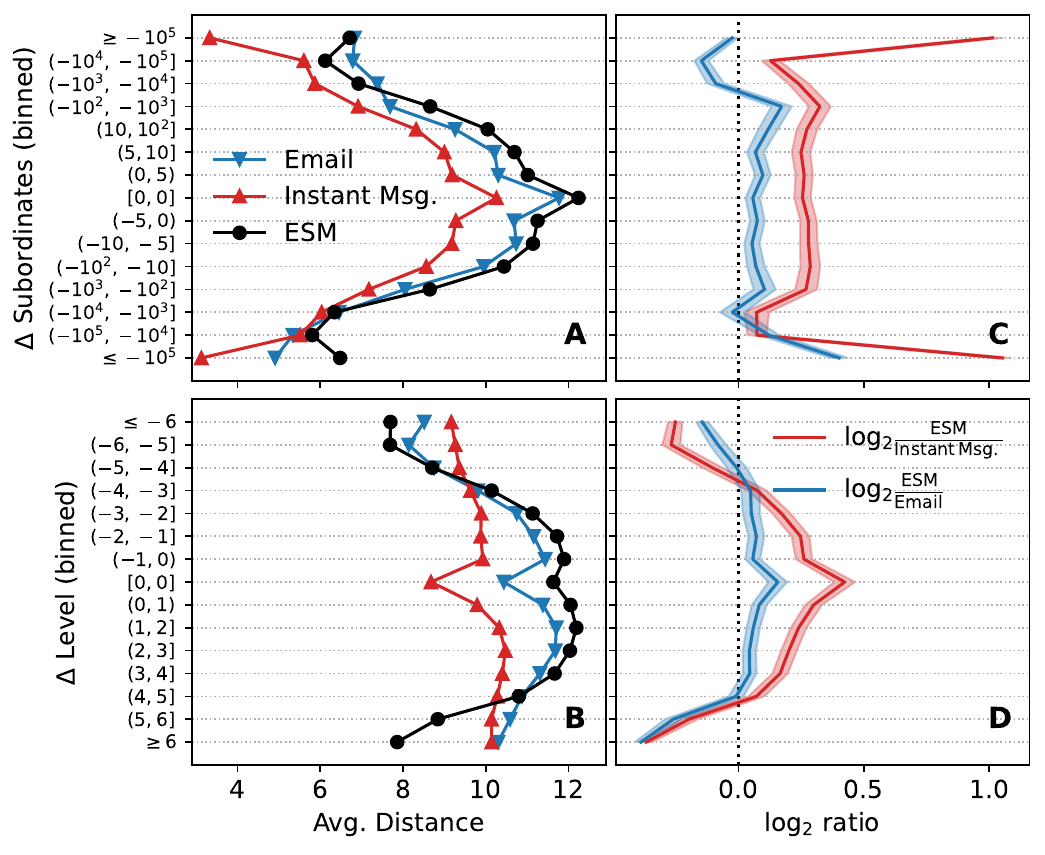}
    \caption{
    We show how ties in Microsoft's communication network bridge the corporate hierarchy horizontally.
    In A--B, for each tie in Microsoft's communication network, we show the relationship between $\text{distance}$ and $\Delta\text{Subordinates}$ (in A) and $\text{Distance}$ and $\Delta\text{Level}$ (in B). 
    For ties that bridge the company vertically according to either of the metrics (on the $y$-axis), we show how distant employees are in the corporate hierarchy (on the $x$-axis).
    In C--D, we again show, for each bin in the $y$-axis, the $log_2$ ratio between the average distance in ESMPs ties and the average distance of ties in email (blue) and instant messaging (red).
    We plot 95\% bootstrapped confidence intervals for the  $log_2$ ratios.
    }
    \label{fig:msft2}
\end{figure}

\vspace{.5mm}
\noindent
\textbf{``Horizontal'' ties across the corporate hierarchy.}
The \textit{distance} between two individuals in the company's hierarchy captures ``vertical'' distance,
e.g., the distance between someone and their manager is 1, the same as the difference in Level ($\Delta\text{Level}$).
But also, if the two individuals are embedded in a small or a big `subtree' of the corporate hierarchy, what we call ``horizontal'' distance. 
For instance, two employees at the bottom of the corporate ladder have the same level, but they may report to the same manager (horizontally close), or they may be in entirely different organizations, having the CEO as their closest common ancestor (horizontally far).
We isolate the notion of ``horizontal'' distance' controlling for the difference in influence (vertical distance;   $\Delta\text{Level}$ and $\Delta\text{Subordinates}$).

In Fig.~\ref{fig:msft2}A--B, we show, for each communication technology, the average distance between ties with different values of $\Delta\text{Level}$ and  $\Delta\text{Subordinates}$. 
Again, to gain a relative sense of magnitude, we estimate the $\log_2$ ratio between the average distance in ESMPs and other communications for ties with different $\Delta\text{Level}$ and $\Delta\text{Subordinate}$ (Fig.~\ref{fig:msft2}C--D).
In general, ties in ESMPs connect employees who are more distant in the corporate ladder, e.g., for ties where $\Delta\text{Level} = 0$, we find that the average distance on ESMPs is 11.6, vs. 8.7, on Instant Messaging and 10.5 on Email. These correspond to statistically significant, positive $\log_2$ ratios.

\vspace{.5mm}
\noindent
\textbf{Centrality across the corporate hierarchy}
Centrality can measure influence within communication networks, and one's level and number of subordinates measure their influence within the corporate hierarchy. 
We study how different communication technologies distribute influence across the corporate hierarchy. 
We calculate the PageRank centrality of each employee, considering the ties associated with each communication technology, i.e., we run PageRank considering \textit{only} the ties associated with ESMPs, then only Email, then only Instant Messaging.
Then, estimate the rank correlation between the distribution of PageRank centrality and the distributions of $\text{Level}$ and $\#\text{Subordinates}$. We find that PageRank centrality in the ESMPs communication network is not (rank) correlated to level $\rho_\text{lev} = -0.02$ and number of subordinates $\rho_\text{sub.} = 0.01$. 
This is not the case for the communication networks of Email and Instant Messaging, where correlations are stronger and statistically significant; 
$\rho_\text{lev} = -0.21$ and  $\rho_\text{sub.} = 0.27$ for Email.
$\rho_\text{lev} = -0.14$ and  $\rho_\text{sub.} =  0.25$ for Instant Messaging.

\begin{figure}
    \centering
    \includegraphics[width=\linewidth]{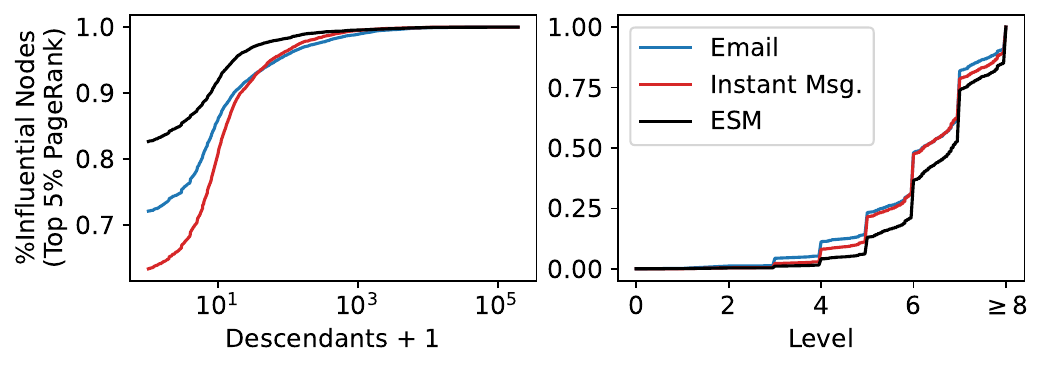}
    \caption{
    Empirical Cumulative Distribution Functions (CDFs) depicting how employees that are influential in the communication network (top 5\% PageRank) are distributed through metrics capturing influence in the company's hierarchy (Descendants, left;  Level, right). 
    Recall that the CDF maps every value in the $x$-axis to the percentage of values in a sample that are smaller or equal to $x$ (in
the $y$-axis). 
    }
    \vspace{-3mm}
    \label{fig:msft3}
\end{figure}

We further study where the most influential nodes in the communication networks are in the corporate hierarchy.
Considering the separate networks of each communication technology, we estimate the PageRank centrality and obtain the top 5\% users with the highest associated PageRank.
Then, we study how these users are distributed across metrics, capturing their influence in the company's hierarchy (Level and Descendants); see  Fig.~\ref{fig:msft3}.
We find that more influential nodes are better distributed in the company's hierarchy in ESMPs than in Email or Instant Messaging,
e.g., 83\% of most influential nodes "communication-wise" have zero descendants in ESM, versus  72\% and 63\% for email and Instant Messaging, respectively.
This result holds if we consider the top 1\% or 10\% users with the highest associated PageRank.
Overall, these results suggest that ESMPs ``democratize'' influence within a company's communication network since highly central employees seem to be better distributed across the company's hierarchy in ESMPs than in other communication technologies.




\section{Discussion}

We study the role of Enterprise Social Media Platforms (ESMPs) in shaping corporate communication networks. 
Through two studies using the full spectrum of corporate communication networks, we find evidence supporting the hypotheses proposed by ESMPs' pioneers~\cite{dimicco2008motivations,brzozowski2009watercooler}.
First, through a robust difference-in-differences analysis across 99 companies, we find that ESMPs act as a `social lubricant,' increasing the number of ties and `bridges' within organizations' communication networks.
Second, through a case study of Microsoft's communication ecosystem, we find that ESMPs increase the flow of information throughout the company, connecting employees far from each other horizontally (across sectors) and vertically (across the company's hierarchy). 
Altogether, our studies suggest that ESMPs boost information flow within companies and increase employees' attention to what is happening outside their immediate working group by adding new, novel ties that bridge either different sectors of the company or different parts of the corporate hierarchy. ESMPs add these ties over and above the already present email and instant messaging networks.
In an ever-changing world where remote work brings new challenges to companies and employees alike~\cite{yu2023large,brown2023effects,yang2022effects}, ESMPs may help create and maintain social structures that foster productivity and well-being.

A key challenge in studying the causal impact of Enterprise Social Media Platforms (ESMPs) like Engage is the difficulty in isolating the causal effect of ESMP adoption.
We circumvent this challenge in this study by considering the \textit{company} to be the unit of analysis rather than trying to compare employees who adopt ESMPs with those who do not.
Nonetheless, our sample still consists of companies that decided to adopt ESMPs, which may be systematically different from companies that do not choose to do so. While this is not a limitation to identifying the causal effect, it can hinder the generalizability of our results.
Effects obtained may be mediated by characteristics of the sample we consider here, and future work could try to address this limitation by conducting experiments or observational analyses with a representative sample of companies.
Another threat to the causal interpretation of the findings in the first study is that a third event might have triggered both the sudden adoption of ESMPs \textit{and} changes to the company's internal communication practices. We argue that this is unlikely to have interfered with the results. Once a company chooses to use VE, it takes a few months to set it up, such that this third event would also be present in the pre-intervention period.

\begin{figure*}
    \centering
    \includegraphics[width=0.9\linewidth]{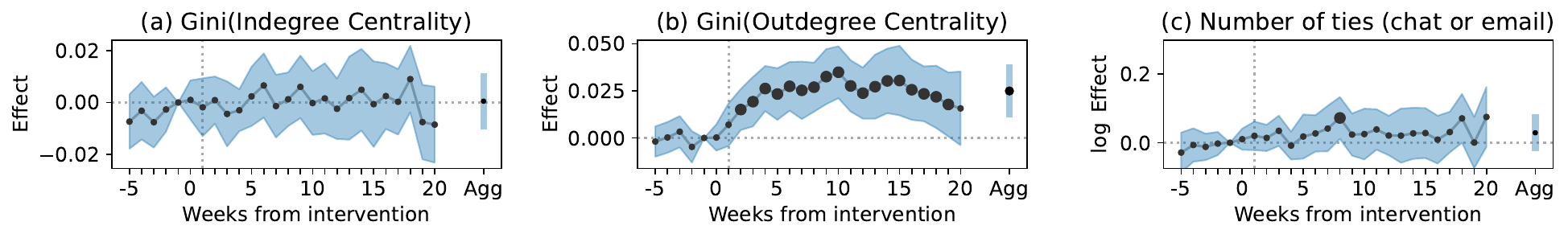}
    \caption{Additional results of our difference in difference analysis. We reproduce Figure~\ref{fig:dd1} for three additional metrics. (A-B) The Gini coefficient is calculated over in- and outdegree centralities of all nodes across all networks. (C) The number of ties calculated only on Instant Messaging and Email.
    }
    \label{fig:app1}
\end{figure*}

Despite our quasi-experimental design and novel data, our work is not without its limitations.
First, it remains unclear how social capital obtained through ESMPs is materialized into real benefits for employees and companies. Do they increase their well-being at work? Future studies could obtain better answers to these questions by employing incentive designs to encourage employees to use ESMPs more often. 
Second, we don't explore the nature of ties. Ties across different communication networks are bound to signify different things, and a threat to the validity of our results is that ESMP ties may be weaker or bring fewer benefits than those in Email and Instant Messaging.
We find that ties on email in email and instant messaging also increased post-ESMP adoption (see Fig~\ref{fig:app1}, which decreases the concerns associated with this limitation. Nonetheless, the precise semantics of ties warrants further investigation.
Third and last, we note that companies may use ESMPs differently. Here, results reflect how the (anonymous) companies (first study) and Microsoft (second study) use ESMPs. Future work could try to disentangle how different strategies of adopting and encouraging ESMP use lead to different outcomes.

\bibliographystyle{ACM-Reference-Format}
\bibliography{references}

\begin{table}[t]
    \centering
    \footnotesize
\caption{Effect of ESMP adoption estimated in our difference in differences regression.}
\begin{tabular}{lrrrr}
\toprule
{} &  Estimate &  Std. Error &  t value &  Pr(>|t|) \\
                               &           &             &          &           \\
\midrule
Number of Ties                 &     0.075 &       0.026 &    2.894 &     0.005 \\
Distinct ties                  &     0.117 &       0.029 &    4.077 &     0.000 \\
Pre-existing ties              &     0.017 &       0.016 &    1.059 &     0.292 \\
One-to-many ties               &     0.133 &       0.032 &    4.226 &     0.000 \\
One-to-one ties                &     0.016 &       0.015 &    1.090 &     0.278 \\
Nbr. of Ties (Chat/Email) &     0.029 &       0.023 &    1.286 &     0.201 \\
Reciprocal Ties                &     0.061 &       0.023 &    2.614 &     0.010 \\
Embeddedness      &    -0.035 &       0.012 &   -2.961 &     0.004 \\
Dispersion        &     0.154 &       0.061 &    2.514 &     0.014 \\
Gini(Degree Centrality)        &     0.013 &       0.005 &    2.709 &     0.008 \\
Gini(Outdegree Centrality)     &     0.025 &       0.006 &    3.959 &     0.000 \\
Gini(Indegree Centrality)      &     0.000 &       0.005 &    0.101 &     0.920 \\
Gini(Page Rank)                &     0.026 &       0.005 &    4.723 &     0.000 \\
\bottomrule
\end{tabular}

    \label{tab:app}
\end{table}

\end{document}